# Frequency agile microwave photonic notch filter with anomalously-high stopband rejection

David Marpaung,* Blair Morrison, Ravi Pant, and Benjamin J. Eggleton

*Centre for Ultrahigh bandwidth Devices for Optical Systems (CUDOS), the Institute of Photonics and Optical Sciences (IPOS), School of Physics, University of Sydney, NSW 2006, Australia*
*Corresponding author: d.marpaung@physics.usyd.edu.au*

We report a novel class microwave photonic (MWP) notch filter with a very narrow isolation bandwidth (10 MHz), an ultrahigh stopband rejection (> 60 dB), a wide frequency tuning (1-30 GHz), and flexible bandwidth reconfigurability (10-65 MHz). This record performance is enabled by a new concept of sidebands amplitude and phase controls using an electro-optic modulator and an optical filter. This new concept enables energy efficient operation in active MWP notch filters, and opens up the pathway to enable low-power nanophotonic devices as high performance RF filters.

Interference mitigation is crucial in modern radiofrequency (RF) communications systems with dynamically changing operating frequencies, such as cognitive radios [1], or for wideband systems such as ultrawideband (UWB) radio [2] and modern radar. To protect sensitive RF receivers in these systems, frequency agile RF filters that can remove interferers or jammers with large variations in frequency, power, and bandwidth are critically sought for. Unfortunately, an RF band-stop or notch filter that can simultaneously provide high resolution, high peak attenuation, large frequency tuning, and bandwidth reconfigurability does not presently exist. State-of-the-art RF notch filters [3-5] are capable of a high peak attenuation (>50 dB) and high selectivity (<10 MHz 3-dB isolation bandwidth measured from the passband) but have limited notch frequency tuning range, in the order of 1.4 GHz [5].

Microwave photonic (MWP) notch filters, on the other hand, are capable of tens of gigahertz tuning and have advanced in terms of performance. These filters can generally be implemented in two ways: via multi-tap filter approach [6-9] where multiple replicas of RF modulated signals are delayed, weighted and combined, to generate nulls in the RF frequency response, or via sideband filtering using an optical filter (OF) [10-14]. Multi-tap approach can generate a notch filter with a narrow bandwidth and relatively high rejection, but exhibits a periodic transfer function which is undesirable for wideband applications. Particular from this filter class, a very narrow notch response (~15 MHz) with relatively wide passband has been demonstrated in [7], but without any tuning capability. The sideband filtering approach requires a high-Q optical resonance to generate the RF notch filter response. Optical cavities such as silicon ring resonators [10, 11], or LiTaO$_3$ WGM resonator [12] have been considered, with isolation bandwidths ranging from 6 GHz [11] down to 10 MHz [12]. But these filters are limited in stopband rejection (~45 dB) due to the challenge in fabrication of the resonators. Recently, RF notch filters based on stimulated Brillouin scattering (SBS) effect in optical fiber [13], or in a photonic chip [14], have been demonstrated, showing narrow isolation bandwidths, in the order of 100 MHz, and multiple-GHz frequency tuning. Nevertheless, to achieve similar performance to state-of-the-art RF filters in terms of isolation bandwidth and rejection is still challenging with these approaches.

In this paper, we demonstrate, for the first time, an MWP notch filter that exhibits a single notch over its entire frequency range, with a high selectivity (~10 MHz) achieved simultaneously with an ultrahigh stopband rejection (>60 dB), a wide continuous frequency tuning (1-30 GHz), and flexible bandwidth reconfigurability (10-65 MHz). This overall record performance is enabled by a new approach in microwave photonics signal processing that allows the use an optical filter (OF) with a very shallow notch, or even a bandpass response, to create an RF filter that exhibits

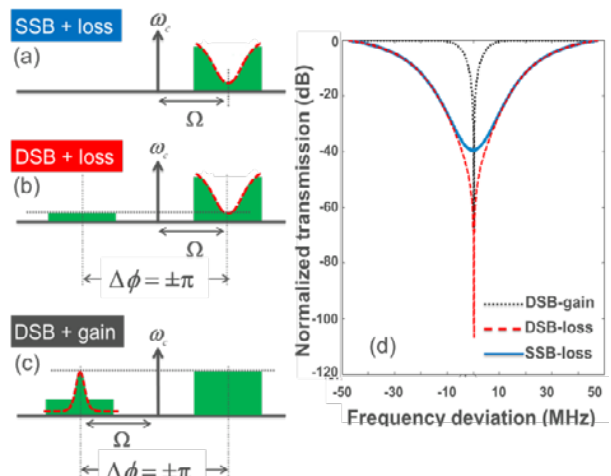

Fig. 1. Principle of operation of notch filter. (a) Conventional single sideband (SSB) scheme, (b) novel scheme using dual sideband (DSB) modulation utilizing loss, (c) novel scheme utilizing gain. (d) Simulated filter responses for the three architectures using SBS (gain = 20 dB) as optical filter.

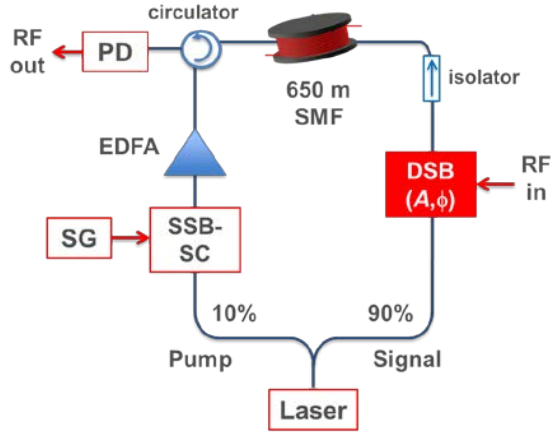

Fig. 2. Experimental setup. SG=RF signal generator, PD=photodetector, SSB-SC= single sideband suppressed carrier modulator, DSB($A,\phi$) = double sideband modulator with phase and amplitude control.

anomalously high stopband rejection.

The novelty of this filter concept lies in the generation and the processing of the optical sidebands. Instead of processing single sideband (SSB) (Fig. 1a) or conventional double-sideband (DSB) signals, i.e., equal amplitude and equal phase (intensity modulation) or equal amplitude and opposite phase (phase modulation), as in traditional MWP filters, here we generate dual sidebands signal with tunable amplitudes and phases. The RF signal is encoded in the optical sidebands with unequal amplitudes and a phase difference, $\Delta\phi$, where $0 < $ modulo $(\Delta\phi) < \pi$, using an electro-optic modulator (EOM). We then exploit both amplitude and phase responses of an OF to equalize these sidebands amplitudes and to produce an anti-phase relation between them ($\Delta\phi = \pm\pi$), only in selected frequency region within the OF response (Fig. 1b). Upon photodetection, the beat signals generated from the mixing of the optical carrier and the two sidebands perfectly cancel at a specific microwave frequency, forming a notch with an anomalously high stopband rejection. This concept is in fact the first photonic implementation of the "bridged-T" notch filter concept in microwaves proposed by Hendrik Bode about eighty years ago [15], where he exploited the concepts of attenuation balance and phase cancellation in a lumped-element circuit to achieve notch filter with quasi-infinite rejection [16].

The novel concept leads to two key advantages; first, the microwave filter peak attenuation is not limited anymore by the OF peak attenuation because the notch is formed by signal cancellation. As will be shown later, this will lead to energy-efficient operation when an active OF is used. Second, the OF is not restricted anymore to produce a notch response, but instead can have a band-pass/gain response, as shown in Fig. 1c. Instead of attenuating the stronger sideband with loss from the OF (Fig. 1b), one can amplify the weaker sideband with gain, to achieve the amplitude equalization required for signal cancellation. Stimulated Brillouin scattering (SBS) gain is an ideal candidate for such an active OF, and it has been long considered for MWP signal processing, due to the fine spectral width of its gain spectrum. For example, a tunable bandpass filter with 23 MHz 3-dB bandwidth was recently reported, exploiting the on-chip SBS gain [17]. The new concept will enable a new direction to exploit the SBS gain spectrum as a high performance notch filter instead.

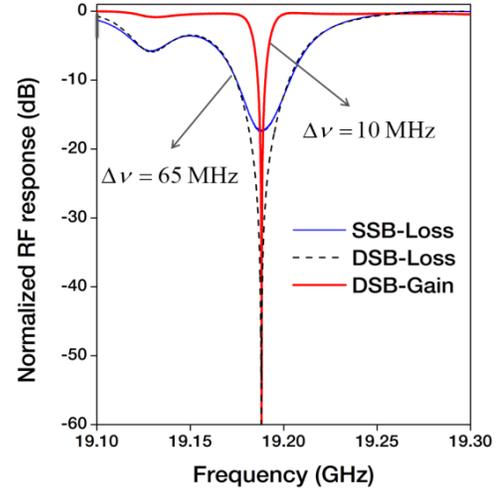

Fig. 3. Experimental results of MWP notch filter responses for SBS gain of ~ 20 dB. DSB-Gain filter exhibit nearly 7 fold isolation bandwidth reduction and > 43 dB rejection enhancement relative to conventional SSB-Loss filter.

In this case, using SBS gain spectrum (Stokes) instead of the SBS loss spectrum (anti-Stokes) is advantageous because for a given $G = g_B I_P L$ parameter [18], where $g_B$ is the Brillouin gain coefficient, $I_P$ is the pump intensity and $L$ is the device length, the gain spectrum exhibit lower 3-dB width compared to the loss spectrum [19]. To illustrate this, we simulate and compare filter responses for the conventional case where the notch filter is generated by SSB modulation and SBS loss spectrum (SSB-Loss), with the novel concept using SBS loss (DSB-Loss), and SBS gain (DSB-Gain). For all three cases, we kept same SBS gain of 20 dB and SBS linewidth of 30 MHz. The results are shown in Fig. 1d. Both the DSB-Loss and DSB-Gain cases show dramatic peak rejection enhancement of >30 dB relative to the SSB-Loss case. Moreover, the DSB-Gain filter shows a bandwidth reduction of nearly 7 fold (10 MHz vs. 69 MHz) compared to SSB-Loss and DSB-Loss filters.

We experimentally demonstrate the novel RF notch filter concept using a setup as shown in the schematic in Fig. 2. Light from a DFB laser (Teraxion Pure Spectrum at $\lambda_L$=1550 nm) was split into pump and signal arms. The pump arm consisted of a single-sideband suppressed carrier (SSB-SC) modulator driven by an RF tone from a signal generator (SG) to generate a pump with tunable frequency [13, 14]. In the signal arm, a double sideband modulator capable of phase and amplitude control (DSB($A,\phi$)) was used to generate first order sidebands from the input RF signal. The pump and the signal were then launched into a 650 m of standard single mode fiber (SMF). The filtered DSB signal was then detected at a high speed photodetector (PD, u2t XPDV2120). The RF filter response was measured using a vector network analyzer (VNA). To implement the (DSB($A$,$\phi$)) we use a dual-parallel Mach-Zehnder modulator (Covega Mach 40-086), driven via a quadrature hybrid coupler (1.7-36 GHz, Krytar). By controlling the three bias voltages of this modulator, the amplitudes and phases of the optical carrier and the RF modulation sidebands can be altered by different amounts [20]. We supply the biases using a multi-channel programmable power supply with 1 mV accuracy (Hameg HM7044G).

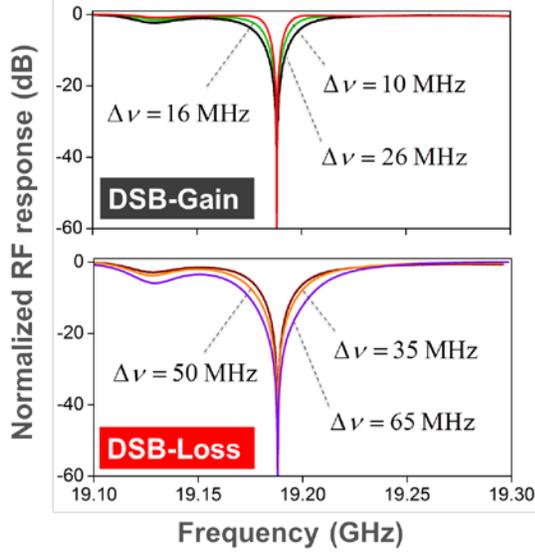

Fig. 4. Experimental results of the MWP notch filter bandwidth tuning achieved by means of tuning the SBS pump power.

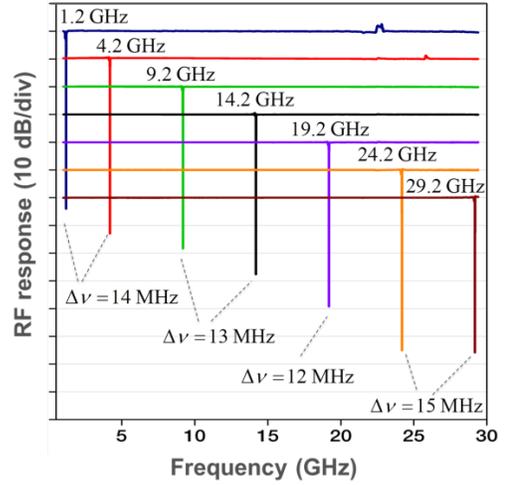

Fig. 5. Experimental results of ultra-wide frequency tuning of the MWP notch filter, preserving a narrow isolation bandwidth of 13±2 MHz and ultrahigh rejection of 60±2 dB.

The dramatic improvement in the filter response using the novel technique is shown in Fig. 3. We measured and compare three different responses, namely the SSB-Loss (solid blue line), the DSB-Loss (dashed black line) and DSB-Gain (thick red line). For all responses the pump power was set at 35.5 mW, corresponding to SBS gain factor of ~20 dB. The SSB-Loss filter exhibit a peak attenuation of 17 dB and a 3-dB width of 65 MHz. The extra peak observed in the experiment was due to multiple SBS modes in the SMF. Switching to DSB-Loss scheme and optimizing the DPMZ biases will improve the peak attenuation of the filter to 61 dB. The most dramatic improvement was obtained when the SBS gain spectrum was used (DSB-Gain), where a record-narrow bandwidth of 10 MHz was obtained simultaneously with a peak attenuation of > 60 dB, which is > 43 dB enhancement relative to the SSB-Loss case.

The frequency agility of the MWP notch filter was demonstrated by means of tuning the bandwidth and the notch frequency. The bandwidth can be reconfigured via varying the SBS gain (by tuning the pump power) and adjusting the DPMZ biases. The result is shown in Fig. 4. For the DSB-Gain case, bandwidth reduction can be achieved by means of increasing the pump power. At pump power of 1.8 mW (corresponding to 1.6 dB gain), the notch filter isolation bandwidth was 26 MHz. Reduction to 10 MHz was achieved using the pump power of 35.5 mW. A reversed trend was observed in the DSB-Loss case, where bandwidth broadening was achieved as the pump power was increased from 1.8 mW (35 MHz) to 35.5 mW (65 MHz). The notch frequency tuning in our filter was achieved by means of varying RF tone frequency supplied from the SG (Fig. 2), which subsequently varied the pump frequency. The notch frequency ($f_{notch}$) is related to pump frequency ($f_{pump}$) via the relation $f_{notch} = f_{pump} - \Omega_B$, where $\Omega_B$ = 10.8 GHz is the Brillouin shift of the SMF used in the experiments. We achieved notch frequency tuning from 1 to 30 GHz, with consistent 3-dB bandwidth of 13±2 MHz and peak attenuation of 60±2 dB, as depicted in Fig. 5. To our knowledge, this is the first time for any filter technology that ultrawide frequency tuning with preserved ultra-narrow isolation bandwidth and ultra-high notch rejection is demonstrated.

Besides high filter performance, the novel technique also leads to energy efficient operation. Since notch rejection is achieved by signal cancellation, significantly lower gain can be used to achieve an ultra-high peak rejection. For example, the DSB-Gain filter at 1.8 mW pump power (1.6 dB gain) requires 58.4 dB lower gain and 55 times lower pump power relative to the SSB-Loss case to achieve the same peak attenuation of 60 dB. Thus, our technique significantly eases the creation of high-performance RF notch filters, and potentially paves the way of enabling low power nanophotonic devices as highly compact RF filters. As an example such a device is the recently reported nanoscale hybrid silicon device which exhibited an SBS gain of approximately 1 dB [21]. We believe, this will lead to a significant advancement in the emerging field of integrated microwave photonics [22].

In conclusions, a novel concept in sideband processing has been proposed and implemented using SBS gain spectrum to achieve an RF photonic notch filter with overall record performance. The concept is applicable to a wide range of optical filters, for example integrated ring resonators, and significantly eases the creation of a high performance RF band-stop filters.

This work was funded by the Australian Research Council (ARC) through its Center of Excellence CUDOS (Grant number CE110001018), and Laureate Fellowship (FL120100029).